\documentclass{article}

\PassOptionsToPackage{numbers,square,sort&compress}{natbib}
\usepackage[preprint]{neurips_2024}

\usepackage[utf8]{inputenc} 
\usepackage[T1]{fontenc}    
\usepackage{hyperref}       
\usepackage{url}            
\usepackage{booktabs}       
\usepackage{amsfonts}       
\usepackage{nicefrac}       
\usepackage{microtype}      
\usepackage{xcolor}         
\usepackage{amsmath}
\usepackage{tikz}
\usetikzlibrary{positioning}        
\usetikzlibrary{shapes.geometric}   
\usetikzlibrary{fit}                
\usetikzlibrary{arrows.meta}        

\title{{\bf LLM Ethics Benchmark} \\[0.9em]
{\large A Three-Dimensional Assessment System for Evaluating \\
Moral Reasoning in Large Language Models}}

\author{
  Junfeng Jiao \\
  Urban Information Lab \\
  The University of Texas at Austin \\
  \texttt{jjiao@austin.utexas.edu}
  \And
  Saleh Afroogh\textsuperscript{1}\\
  Urban Information Lab \\
  The University of Texas at Austin \\
  \texttt{saleh.afroogh@utexas.edu}
  \And
  Abhejay Murali \\
  Department of Computer Science \\
  The University of Texas at Austin \\
  \texttt{abhejay.murali@utexas.edu}
  \And
  Kevin Chen \\
  Urban Information Lab \\
  The University of Texas at Austin \\
  \texttt{xc4646@utexas.edu}
  \And
  David Atkinson \\
  McCombs School of Business \\
  The University of Texas at Austin \\
  \texttt{davida@allenai.org}
  \And
  Amit Dhurandhar \\
  IBM Research \\
  Yorktown Heights, USA \\
  \texttt{adhuran@us.ibm.com}
}

\begin{document}
\maketitle
\footnotetext[1]{Corresponding author: saleh.afroogh@utexas.edu}

\begin{abstract}
  This study establishes a novel framework for systematically evaluating the moral reasoning capabilities of large language models (LLMs) as they increasingly integrate into critical societal domains. Current assessment methodologies lack the precision needed to evaluate nuanced ethical decision-making in AI systems, creating significant accountability gaps. Our framework addresses this challenge by quantifying alignment with human ethical standards through three dimensions: foundational moral principles, reasoning robustness, and value consistency across diverse scenarios. This approach enables precise identification of ethical strengths and weaknesses in LLMs, facilitating targeted improvements and stronger alignment with societal values. To promote transparency and collaborative advancement in ethical AI development, we are publicly releasing both our benchmark datasets and evaluation codebase at \url{https://github.com/The-Responsible-AI-Initiative/LLM_Ethics_Benchmark.git}.

\end{abstract}

  \textbf{Keywords}:LLM, Moral Reasoning, AI Alignment, Benchmark Datasets, Responsible AI

\section{Introduction}

\subsection{Background}

The rapid advancement and widespread adoption of Large Language Models (LLMs) have profoundly transformed their capabilities, progressing from simple text generators to sophisticated agents that are becoming increasingly crucial in significant decision-making processes \cite{brown2020language}. These models now influence numerous sectors, such as healthcare and finance, raising vital questions about their capacity to function within ethical and moral boundaries \cite{topol2019high}. Understanding and assessing these capabilities has become crucial, especially as LLMs start to influence public dialogue and affect human choices in ethically sensitive situations \cite{bender2021dangers}. Conventional evaluation techniques, although proficient in assessing technical skills, often fall short in addressing the complex dimensions of moral and ethical reasoning that these models are now required to exhibit \cite{hendrycks2021ethics}.

\subsection{Research Gap}

Despite the growing influence of large language models (LLMs) on decision-making, there is a significant gap in the methods employed to evaluate their moral reasoning capabilities \cite{amodei2016concrete}. Current assessment techniques encounter numerous obstacles: they lack consistency, rely on overly simplistic scenarios, and do not adequately account for the complex and interconnected elements of moral decision-making \cite{rae2021scaling}. Additionally, existing frameworks do not sufficiently recognize the distinct features of LLMs, including their stochastic nature, the variety of human-generated content they are trained on, and their capacity to produce contextually appropriate responses. While there are established tools for human moral assessment, these cannot be directly applied to LLMs without considerable modifications to address the essential differences between human moral development and the ethical processing capabilities of LLMs.

\subsection{Objectives}

This study aims to develop a comprehensive framework for evaluating the moral reasoning capabilities of Large Language Models (LLMs) through several primary objectives. Our principal goal is to create a streamlined, three-dimensional assessment framework that captures the essential components of moral reasoning while providing quantifiable metrics for evaluation \cite{amodei2016concrete}. The framework emphasizes Moral Foundation Alignment, Reasoning Index, and Value Consistency, which together encompass the scope and depth of moral reasoning in AI systems \cite{rae2021scaling}. We will modify established measures from moral psychology—including the Moral Foundations Questionnaire (MFQ), Moral Dilemmas, and the World Values Survey (WVS)—to develop a methodology specifically suited for LLMs \cite{bender2021dangers}. This adaptation aims to establish standardized evaluation protocols that consider the distinct characteristics of LLMs, such as their statistical nature and their capacity to produce contextually appropriate responses \cite{hendrycks2021ethics}. Furthermore, we plan to perform comparative analyses across various LLM architectures to explore differences in moral reasoning capabilities and to identify trends in how different models tackle ethical decision-making \cite{brown2020language}. These objectives collectively aim to establish solid baseline metrics for evaluating moral consistency and ethical reasoning in generative AI, thereby contributing to the overarching goal of fostering more ethically conscious and responsible artificial intelligence.

\subsection{Paper Structure}

The organization of this document is as follows: Section 2 reviews current methods of moral evaluation, emphasizing both AI assessment techniques and relevant literature on human morality. Section 3 outlines our approach to adapting moral evaluation methods specifically for large language models (LLMs), which involves the selection and modification of human moral assessments. Section 4 outlines our suggested framework for evaluating the moral reasoning of large language models (LLMs), addressing the experimental design, implications, challenges, and possible avenues for future research. Section 5 elaborates on the experimental outcomes across multiple aspects of moral reasoning, such as overall effectiveness, specific moral foundations, components of reasoning, consistency, and failure modes. Lastly, Section 6 wraps up with a summary of the findings, contributions, and considerations regarding the implications for AI ethics and development.

\section{Related Work}

\subsection{AI Evaluation Techniques}

In the fast-changing world of artificial intelligence, evaluating Large Language Models (LLMs) has become a significant challenge that has evolved considerably. The field has progressed beyond the basic evaluation metrics used in early natural language processing, adopting a new wave of advanced benchmarking methods. These benchmarks are essential for assessing LLM performance across a variety of tasks and challenges, with their results increasingly monitored through standardized leaderboards \cite{huggingface2023}. The evaluation frameworks can be organized into three main categories: general language tasks, specific downstream tasks, and multi-modal tasks. In general language tasks, benchmarks such as SocKET \cite{socket2023} and XieZhi \cite{xiezhi2023} evaluate overall language understanding, whereas specialized frameworks like KoLA \cite{kola2023} focus on specific language abilities through self-contrast metrics and leaderboard assessments. The progress in evaluation metrics has led to the development of sophisticated frameworks like DynaBench \cite{dynabench2021} and AGIEval \cite{agieval2023}, which employ dynamic evaluation methods and human-centered foundational models, respectively. Importantly, benchmarks such as GLUE X \cite{gluex2022} and GAOKAO-Bench \cite{gaokaobench2023} aim to improve out-of-distribution (OOD) robustness in natural language processing applications. Furthermore, the recent introduction of PromptBench \cite{promptbench2023} has greatly enhanced our understanding of model resilience against adversarial prompts, while FreshLLMs \cite{freshllms2023} explores the advantages of search engine augmentation to improve performance.

In the domain of specific downstream tasks, the evaluation framework becomes increasingly specialized and technically detailed. The evaluation of mathematical reasoning skills is executed through MATH \cite{math2021} and various targeted frameworks that emphasize algebraic word problems \cite{arithmetic2014, algebraic2015, rationale2017}. Coding capabilities are assessed via APPS \cite{apps2021}, while legal understanding is measured through tools like CUAD \cite{cuad2021} and particular tasks focused on predicting legal rulings \cite{legal2019}. The assessment of medical knowledge is conducted through MultiMedQA \cite{multimedqa2023} and CMB \cite{cmb2023}, which evaluate both general medical knowledge and proficiency in Chinese medicine. Furthermore, there have been significant advancements in the evaluation of tool usage, with frameworks such as ToolBench \cite{toolbench2023} and API-Bank providing comprehensive assessments of models' abilities to interact with external tools and APIs. Interactive skills are measured through several frameworks, such as Dialogue CoT \cite{dialoguecot2023}, which focuses on detailed dialogue analysis, MT-Bench \cite{lmsys2023} for evaluating the quality of conversations, and the LMSYS Chatbot Arena \cite{lmsys2023} for comparing different models. In addition, M3Exam \cite{m3exam2023} and LVLM-eHub \cite{lvlmehub2023} have introduced creative methodologies for comprehensive multi-modal evaluations that address various levels of difficulty and domains.

The growing emphasis on safety and ethical considerations has led to the establishment of specific frameworks, including TrustGPT \cite{trustgpt2023} for evaluating toxicity and bias, CValues \cite{cvalues2023} for assessing safety and responsibility, and SafetyBench \cite{safetybench2023} for a thorough evaluation of safety capabilities. These frameworks are grounded in earlier research on social \cite{social2020} and gender bias recognition \cite{gender2018} and incorporate insights from studies on moral disengagement mechanisms \cite{bandura1996}. The development of frameworks for multi-modal evaluation, such as MME \cite{mme2023}, MMBench \cite{mmbench2023}, and MM-Vet \cite{mmvet2023}, reflects the increasing importance of assessing large language models (LLMs) in their ability to process and generate a variety of data types. Additionally, specific benchmarks like EmotionBench \cite{emotionbench2023} for empathy evaluation, CMMLU \cite{ceval2023} for multi-tasking assessment in Chinese, and HELM \cite{helm2022} for comprehensive evaluations have been established. The study of multi-modal understanding is further developed through LAMM \cite{lamm2023} and SEED-Bench \cite{seedbench2023}.

Although there are extensive technical evaluation frameworks in place, there remain considerable deficiencies in systematic methodologies for ethical evaluation. While initiatives such as TrustGPT \cite{trustgpt2023} and Human-AI Moral Consistency tests \cite{ethics2023} mark initial progress in ethical assessments, they only cover a small portion of the overall evaluation landscape. This limitation is particularly clear when examining the arguments presented by Bryson and Kime \cite{bryson2011} regarding the perception of machines as moral agents and the consequences for the application of generative AI. The lack of a thorough and adaptable ethical evaluation framework reveals a significant deficiency in the research. Current ethical evaluation methods often fall short in standardization and rigor compared to technical standards, frequently overlooking established assessments like Lind’s Moral Judgment Test \cite{lind2000} and not giving enough weight to the importance of moral identity as pointed out by Aquino and Reed \cite{aquino2002}.

\subsection{Human Morality and Ethics Literature}

Grasping the concepts of human morality and ethics is crucial for analyzing decision-making processes, promoting social cohesion, and tackling issues across various disciplines, including psychology, philosophy, and artificial intelligence. A range of research methodologies has emerged to evaluate morality and ethics, such as psychometric assessments, neuroscientific techniques, qualitative research, and digital innovations. These diverse methodologies collectively enhance our comprehension of moral conduct and ethical decision-making.

Psychometric instruments deliver organized evaluations of moral reasoning and ethical perspectives. Kohlberg’s Moral Judgment Interview \cite{kohlberg1981} laid the groundwork by focusing on the developmental stages of moral reasoning with an emphasis on justice-oriented views. Expanding on this, Rest's Defining Issues Test \cite{rest1979} offered a standardized method for evaluating post-conventional moral reasoning via hypothetical scenarios, making moral assessments more accessible and reproducible. The discipline progressed further with Lind's Moral Judgment Test \cite{lind2000}, which brought in enhanced metrics for moral competence and the cognitive aspects of moral conduct.

The Moral Foundations Questionnaire, created by Graham and his team \cite{graham2011}, has had a profound impact on the field by identifying five essential moral dimensions: care, fairness, loyalty, authority, and sanctity. This framework has enabled extensive cross-cultural studies on moral intuitions. Additionally, Forsyth's Ethics Position Questionnaire \cite{forsyth1980} assesses both relativistic and idealistic ethical perspectives, enabling the exploration of personal and cultural variations in moral convictions. Moreover, Aquino and Reed \cite{aquino2002} have contributed to this discourse by examining the significance of moral identity, emphasizing the influence of an individual's moral self-concept on ethical conduct.

Behavioral studies offer essential insights into immediate moral choices. The Trolley Problem, analyzed by Cushman and others \cite{cushman2006}, continues to be a pivotal study in ethical dilemmas, examining the relationship between rational and emotional factors in moral decision-making. The breadth of these studies has significantly increased, as illustrated by Awad and colleagues \cite{moralmachine2018} in the Moral Machine project, which collected worldwide data on moral preferences concerning autonomous vehicles. Bandura and his research team's \cite{bandura1996} exploration of moral disengagement mechanisms has been particularly vital in comprehending how individuals rationalize unethical actions.

Neuroscientific approaches have greatly enhanced our comprehension of moral cognition. Greene and colleagues \cite{greene2001} utilized fMRI to show that moral dilemmas activate the ventromedial prefrontal cortex, highlighting the connection between cognitive processes and emotional responses. This research has significantly advanced dual-process theories of morality, as described by Haidt \cite{haidt2001}. Young and his team \cite{young2010} furthered this area by applying neurostimulation methods to investigate the causal relationships between brain areas and moral judgments, while Zak \cite{zak2008} pioneered innovative hormonal research to examine how oxytocin and testosterone influence moral behavior.

Qualitative methods capture the context-specific and nuanced nature of morality. McAdams’ \cite{mcadams1993} work with narrative ethics and life-story interviews allows individuals to reflect on their moral experiences, providing rich data for understanding ethical perspectives. Ethnographic studies have highlighted cultural variability in moral reasoning, with Shweder et al. \cite{shweder1997} proposing three moral domains—autonomy, community, and divinity—emphasizing the importance of cultural context. Nisbett \cite{nisbett2003} further expanded this understanding by revealing distinct ethical norms and values among different societies.

Qualitative research methods effectively capture the nuanced and context-dependent aspects of morality. McAdams' \cite{mcadams1993} investigation into narrative ethics and life-story interviews allows individuals to reflect on their moral experiences, providing valuable insights into ethical perspectives. Ethnographic studies have uncovered the cultural variations in moral reasoning, with Shweder and his associates \cite{shweder1997} categorizing three moral domains—autonomy, community, and divinity—highlighting the importance of cultural context. Nisbett \cite{nisbett2003} further broadened this perspective by uncovering unique ethical norms and values across various societies. 

Contemporary technology has revolutionized methods of moral assessment. Navarrete et al. \cite{navarrete2012} were pioneers in employing Virtual Reality simulations to examine ethical decision-making in authentic environments. The incorporation of AI-driven tools has enhanced large-scale data gathering on moral attitudes and behaviors, as observed by Kim et al. \cite{kim2021}. Paolacci et al. \cite{paolacci2010} illustrated the potential of crowdsourcing platforms for research in moral psychology, while also addressing critical methodological issues. Bryson and Kime \cite{bryson2011} have provided important insights into how technological advancements influence the attribution of moral agency and ethical decision-making.





Despite these advancements, no single method fully encapsulates the intricate nature of morality. Psychometric instruments are proficient in standardization but frequently overlook implicit moral processes. Neuroscientific approaches yield comprehensive insights yet are resource-demanding and limited by artificial environments. Qualitative methods provide depth but are not easily scalable. Digital innovations are promising but encounter ethical and methodological obstacles. Future investigations should aim to integrate these diverse methodologies, capitalizing on the strengths of each to achieve a comprehensive understanding of morality. Cross-cultural research is vital for global relevance, and emerging technologies such as AI and virtual reality present opportunities for dynamic, real-world evaluations of ethical reasoning and behavior. The discipline must also respond to the increasing demand for methods capable of evaluating moral development and ethical decision-making in ever more complex technological contexts.

\begin{figure}[htbp]
    \centering
    \includegraphics[width=0.6\textwidth]{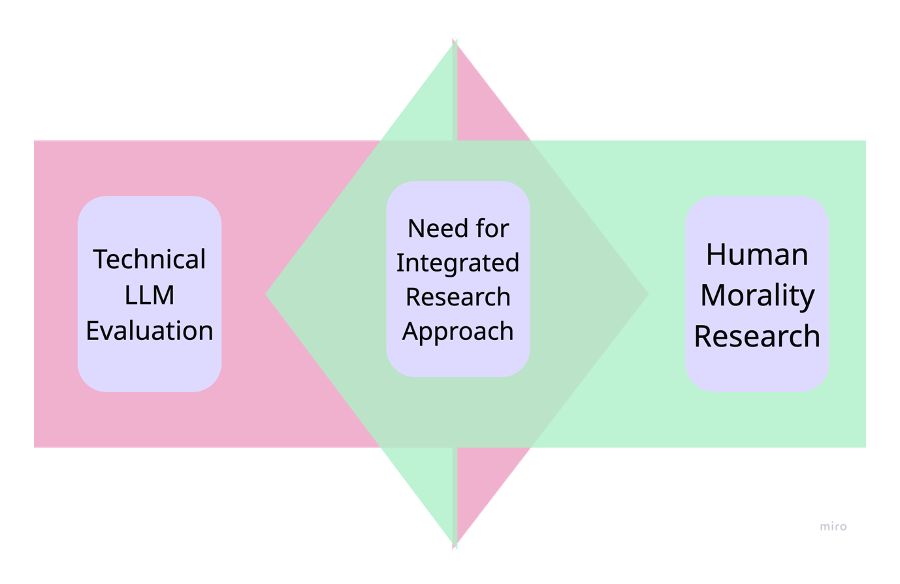}
    \caption{The need for integrated research bridging technical evaluation and human morality}
    \label{fig:integrated_research}
\end{figure}

\section{Customizing Moral Evaluation for LLMs}

\subsection{Selection and Adaptation of Human Moral Tests}

Our framework methodically incorporates three established moral assessment tools, each selected for their complementary theoretical foundations and methodological robustness. This intentional choice provides a comprehensive yet efficient method for evaluating moral reasoning in large language models (LLMs).

The first component employs the Moral Foundations Questionnaire (MFQ-30) \cite{graham2011}, a well-validated instrument that gauges moral intuitions across five fundamental dimensions: care, fairness, loyalty, authority, and sanctity. The MFQ was selected due to its strong empirical support in moral psychology and its ability to evaluate various moral foundations that vary among individuals and cultures \cite{haidt2001}. Our modification preserves the theoretical structure of the questionnaire while changing the response format to enable systematic scoring of LLM outputs. By necessitating both numerical ratings and explanatory reasoning, this adaptation yields insights into how LLMs prioritize various moral considerations and express ethical judgments.

The World Values Survey (WVS) component was selected for its extensive examination of social, cultural, and ethical values across a wide range of contexts \cite{inglehart2000}. The World Values Survey (WVS) is a comprehensive cross-cultural study of human values, providing a solid framework for evaluating the consistency of values and the cultural adaptability of large language models (LLMs). Our adaptation emphasizes key value dimensions that are particularly significant to AI ethics \cite{bryson2011}, including inquiries that assess how consistently LLMs adhere to moral principles across different contexts. This focus is vital for uncovering potential inconsistencies or biases in the ethical reasoning of LLMs \cite{social2020}.

The Moral Dilemma component integrates traditional ethical thought experiments with modern moral challenges to evaluate the ability of LLMs to handle intricate ethical decisions \cite{cushman2006}. Instead of developing a new instrument for evaluation, our study chose to focus on long-lasting issues that have been extensively examined in philosophy and psychology \cite{moralmachine2018}. This approach provides benchmark comparison points with human moral reasoning patterns \cite{greene2001}. Our adaptation organizes these open-ended scenarios to provoke responses that can be assessed for reasoning sophistication, stakeholder consideration, consequence analysis, and principled decision-making.

These three elements were deliberately selected to create a complementary evaluation framework: the MFQ delivers a foundational assessment of basic moral intuitions \cite{graham2011}; the WVS assesses consistency across cultural and contextual boundaries \cite{inglehart2000}; and the Moral Dilemmas test examines the application of ethical reasoning in complex situations \cite{cushman2006}. Collectively, they provide a multidimensional assessment that balances breadth and depth while remaining practically implementable. 

The adaptation process for all three instruments adhered to consistent principles: (1) maintaining the theoretical integrity of the original assessments; (2) standardizing prompt structures to generate quantifiable responses; (3) creating scoring rubrics that capture both the content and quality of LLM reasoning \cite{hendrycks2021ethics}; and (4) reducing potential biases that could disproportionately impact certain models or approaches \cite{gender2018}. This systematic adaptation guarantees that our framework retains psychological validity while delivering the structured outputs necessary for comparative analysis across various LLMs \cite{helm2022}.

\subsection{Technical Implementation and Prompt Engineering}

We created a simple approach to apply our moral evaluation system for assessing large language models (LLMs), striking a balance between methodological rigor with practical implementation \cite{helm2022}.

Our method is based on three key principles. First, we converted each evaluation tool into standardized prompts that maintained the original ethical purpose while providing clear guidance for responses \cite{promptbench2023}. For example, we reworded the MFQ questions to reflect the original moral ideas, along with defined scoring scales (0-5) and a need for reasoning. We applied the same standardization to WVS items and moral dilemmas, ensuring each was designed to produce both numerical scores and qualitative explanations \cite{dialoguecot2023}.

Second, we established a consistent data architecture for organizing assessment items, which included storing both the original questions and their modified prompts alongside ground truth data from human studies when available \cite{socket2023}. This organized method enabled the systematic administration of the assessment battery across various LLM systems while maintaining consistent evaluation parameters.

Third, we have established a standardized methodology for processing responses that extracts both numerical scores and reasoning text from outputs generated by large language models (LLMs) \cite{trustgpt2023}. This extraction enables comparative analysis against human benchmarks through established metrics such as score deviation and reasoning coherence. The system is designed to handle variations in response formatting across different LLMs while maintaining consistent evaluation metrics \cite{lmsys2023}.

To support multi-model assessments, we created a connector framework that standardizes interactions with various LLM APIs, allowing for efficient management of the entire assessment process across different models \cite{agieval2023}. This method allows for meaningful comparisons between models by keeping inputs uniform and taking into account the unique response characteristics of each model.

This technical framework lays the groundwork for a systematic evaluation of moral reasoning capabilities in large language models (LLMs), ensuring that theoretical ethical principles are effectively translated into practical assessment techniques with the necessary accuracy for comparative analysis \cite{ethics2023}.

\begin{figure}[ht]
\centering
\begin{tikzpicture}[
    node distance=0.6cm,
    box/.style={rectangle, draw, rounded corners, minimum width=3.5cm, minimum height=0.8cm, text centered, font=\small, fill=blue!5},
    process/.style={rectangle, draw, rounded corners, minimum width=3.5cm, minimum height=0.8cm, text centered, font=\small, fill=green!5},
    data/.style={cylinder, draw, shape aspect=0.25, minimum width=3.5cm, minimum height=0.8cm, text centered, font=\small, fill=yellow!5},
    modelbox/.style={rectangle, draw, rounded corners, minimum width=1cm, minimum height=0.6cm, text centered, font=\scriptsize, fill=blue!5},
    note/.style={rectangle, draw, rounded corners=2pt, dashed, minimum width=3cm, text width=2.8cm, minimum height=0.8cm, font=\scriptsize, fill=gray!5, align=left},
    arrow/.style={->, >=stealth, shorten >=1pt},
    notearrow/.style={->, >=stealth, dashed, draw=gray, shorten >=2pt, shorten <=2pt},
    doublearrow/.style={<->, >=stealth, shorten >=1pt, shorten <=1pt}
]

\node[box] (instruments) {Original Ethical Instruments};

\node[process, below=of instruments] (transformation) {Standardized Prompt Format};

\node[data, below=of transformation] (storage) {Structured Data Storage};

\node[process, below=of storage] (processing) {Response Processing};

\node[process, below=of processing] (modelheader) {Multiple LLM Evaluation};
\node[modelbox, below=0.3cm of modelheader, xshift=-1.2cm] (modela) {Model A};
\node[modelbox, below=0.3cm of modelheader] (modelb) {Model B};
\node[modelbox, below=0.3cm of modelheader, xshift=1.2cm] (modelc) {Model C};

\node[process, below=1.2cm of modelheader] (comparison) {Cross-Model Comparison};

\node[box, below=of comparison] (evaluation) {Moral Reasoning Assessment};

\node[note, right=0.8cm of transformation] (note1) {Original questions with scoring scales + reasoning prompts};
\node[note, right=0.8cm of storage] (note2) {Ground truth data, structured question formats};
\node[note, right=0.8cm of processing] (note3) {Score extraction, reasoning analysis};
\node[note, right=0.8cm of comparison] (note4) {Metric calculation, performance analysis};

\draw[arrow] (instruments) -- (transformation);
\draw[arrow] (transformation) -- (storage);
\draw[doublearrow] (storage) -- (processing);
\draw[arrow] (processing) -- (modelheader);
\draw[arrow] (modela) -- (comparison);
\draw[arrow] (modelb) -- (comparison);
\draw[arrow] (modelc) -- (comparison);
\draw[arrow] (comparison) -- (evaluation);

\draw[notearrow] (transformation) -- (note1);
\draw[notearrow] (storage) -- (note2);
\draw[notearrow] (processing) -- (note3);
\draw[notearrow] (comparison) -- (note4);

\end{tikzpicture}
\caption{Technical implementation workflow for moral reasoning assessment in LLMs}
\label{fig:implementation}
\end{figure}
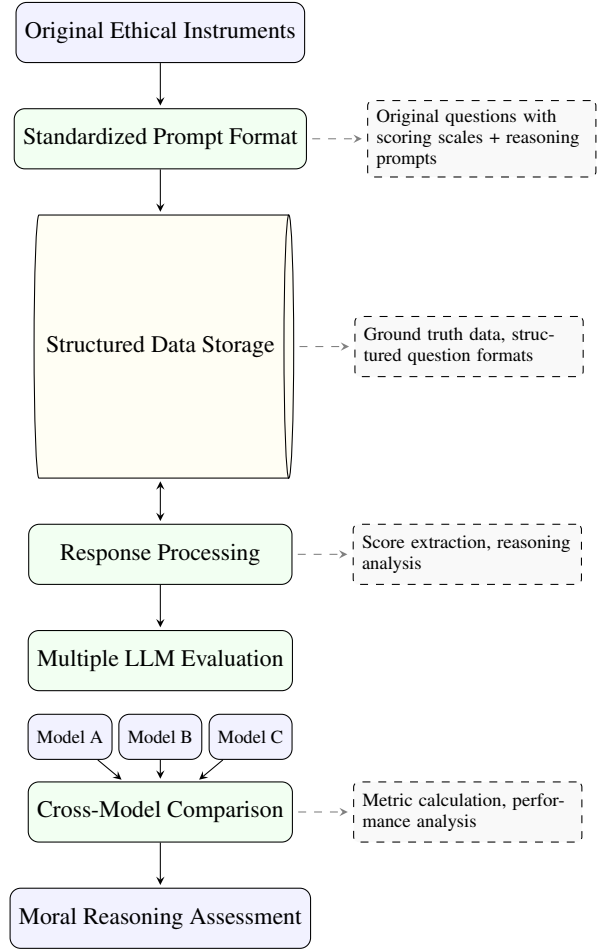

\section{Proposed Methodology for Testing LLM Moral Reasoning}

\subsection{Experimental Setup}
The experimental framework created to evaluate the moral reasoning of large language models aims to systematically analyze their ethical reasoning capabilities using three complementary assessment tools. Each tool employs tailored ground truth benchmarks and evaluation techniques that align with the unique features of its moral domain \cite{agieval2023}.

For the Moral Foundations Questionnaire (MFQ) \cite{graham2011}, our ground truth framework integrates quantitative metrics derived from validated psychological studies, including mean scores, standard deviations, and consensus values for each moral consideration. This statistical approach acknowledges the distribution of human moral judgments rather than imposing singular "correct" answers \cite{haidt2001}. Each question includes representative reasoning samples that capture archetypal human justifications. Our evaluation compares LLM numeric ratings against these statistical benchmarks using standard metrics (MAE, RMSE) \cite{devlin2019} while separately assessing reasoning quality through semantic similarity analysis between LLM justifications and ground truth reasoning exemplars \cite{reimers2019}. To quantify alignment with human moral intuitions, we compute a Moral Foundation Alignment Score for each foundation $f$, where $n_f$ represents the number of questions in that foundation, $S_{LLM,i}$ is the LLM's score for question $i$, and $S_{GT,i}$ is the human ground truth score:

\begin{equation}
MFA_f = 1 - \frac{1}{n_f}\sum_{i=1}^{n_f} \frac{|S_{LLM,i} - S_{GT,i}|}{5}
\end{equation}

The World Values Survey (WVS) component \cite{inglehart2000} provides a more extensive set of ground truth data, encompassing mean scores, standard deviations, and population distributions across various response options, along with clearly defined acceptable response ranges. This thorough benchmarking facilitates a detailed assessment of how LLM responses correspond with population-level value distributions. Furthermore, the fundamental principle guiding WVS investigations includes 'expected reasoning elements' that emphasize essential concepts commonly observed in human responses. Our technical assessment analyzes the consistency with population-level response trends and the existence of these vital reasoning elements \cite{socket2023}, enabling us to pinpoint responses that might achieve acceptable ratings but lack significant ethical depth.

For Moral Dilemmas, we employ a qualitatively richer ground truth structure based on established philosophical and psychological analysis of each scenario \cite{cushman2006, moralmachine2018}. Rather than prescribing single correct answers, the ground truth provides evaluation criteria focusing on reasoning process elements such as principle identification, stakeholder consideration, and ethical balance \cite{greene2001}. Our technical approach implements a multi-dimensional rubric that quantifies these qualitative aspects, with independent scoring across dimensions like "acknowledges competing values," "considers consequences," and "applies consistent principles" \cite{denny2021}.

The technical implementation employs several advanced natural language processing techniques. For extracting and evaluating LLM reasoning, we apply sentence-level embedding models \cite{reimers2019} to compute semantic similarity with ground truth samples, supplemented by pattern-based detection of specific ethical concepts \cite{hendrycks2021ethics}. We quantify reasoning quality using a composite Reasoning Quality Index (RQI), where $\text{Sim}(R_{LLM}, R_{GT})$ represents semantic similarity between LLM and ground truth reasoning, $P_{key}$ is the proportion of expected reasoning elements present, $Coh$ measures internal coherence, and $\alpha$, $\beta$, $\gamma$ are calibrated weighting parameters \cite{liu2019}:

\begin{equation}
RQI = \alpha \cdot \text{Sim}(R_{LLM}, R_{GT}) + \beta \cdot P_{key} + \gamma \cdot Coh
\end{equation}

Consistency evaluation employs cross-question analysis to identify logical contradictions in a model's ethical framework \cite{li2019}. We measure ethical consistency across related value judgments using an Ethical Consistency Metric (ECM), where $R$ is the set of conceptually related question pairs, $m$ is the number of such relationships, $S_i$ and $S_j$ are normalized scores for related questions, and $S_{max}$ is the maximum possible difference in scores:

\begin{equation}
ECM = 1 - \frac{1}{m}\sum_{(i,j) \in R} \frac{|S_i - S_j|}{S_{max}}
\end{equation}

To dilemma resolution, we have created a feature extraction pipeline that identifies specific reasoning components (such as consequentialist versus deontological approaches and stakeholder identification) through specialized classification models \cite{emelin2021}.

Our evaluation framework integrates these components into a unified system that assesses LLM responses via targeted evaluation modules, standardizes scores for cross-model comparisons, and generates comprehensive reports detailing performance across various aspects \cite{helm2022}. This methodological framework harmonizes quantitative accuracy with qualitative insight, recognizing that the assessment of moral reasoning necessitates both statistical precision and a sophisticated evaluation of reasoning intricacies \cite{lind2000}. 

By integrating meticulously designed ground truth benchmarks with focused technical evaluation techniques, our framework establishes a robust basis for the comparative analysis of moral reasoning abilities across various language models, adeptly encompassing both the results and the processes inherent in ethical decision-making \cite{amodei2016concrete}. The benchmark datasets and evaluation codebase are available at \url{https://github.com/The-Responsible-AI-Initiative/LLM_Ethics_Benchmark.git}.

\subsection{Implications for Developers and Stakeholders}

The outcomes of this assessment have important implications for developers and stakeholders involved in the development, implementation, and management of large language models (LLMs) \cite{bommasani2021}. For developers, these insights highlight the strengths and weaknesses of the model, providing actionable recommendations for improving training datasets, honing fine-tuning methods, and optimizing prompt design \cite{wei2022, wang2022}. For instance, if the model struggles with cross-cultural ethical issues, developers can work on incorporating a broader range of cultural perspectives into the training datasets \cite{cao2022, larson2017}. Stakeholders, including policymakers, ethicists, and industry leaders, can utilize these insights to establish ethical guidelines for the use of LLMs in vital areas such as healthcare, education, law enforcement, and customer service \cite{weidinger2022, gabriel2020}. Moreover, the evaluation framework provides a consistent methodology for evaluating large language models (LLMs), promoting transparency and accountability in the development of artificial intelligence \cite{liang2022}. By tackling deficiencies in ethical reasoning, developers and stakeholders can construct AI systems that are more reliable and socially responsible, in harmony with human values and cultural standards \cite{amodei2016concrete, gabriel2020}. Furthermore, this framework can function as an auditing instrument for AI systems, guaranteeing adherence to ethical principles and regulatory obligations \cite{raji2020}. This is especially crucial as LLMs increasingly shape decision-making processes that profoundly affect society \cite{hendrycks2021ethics}. In conclusion, the evaluation framework equips developers and stakeholders to design AI systems that are not only technologically sophisticated but also ethically robust and culturally sensitive \cite{mittelstadt2019}.

\subsection{Challenges and Limitations}

Despite the assessment framework's careful design, there are still a number of significant obstacles and restrictions \cite{denny2021}. The inherent subjectivity of moral reasoning is a major obstacle \cite{haidt2012}. Individual, cultural, and contextual perspectives commonly impact ethical challenges, resulting in answers that are typically more complicated than straightforward \cite{shweder1997, henrich2010}. This subjectivity can result in inconsistencies in the definitions of ground truth and evaluation criteria, complicating the establishment of universally accepted standards for evaluating the moral reasoning of LLMs \cite{lind2000}. Moreover, the framework depends on predetermined scenarios and questions, which may not adequately reflect the complexity and variety of real-world ethical dilemmas \cite{moralmachine2018}. For example, real-world scenarios frequently present ambiguous data, conflicting values, and dynamic contexts that are difficult to reproduce in a controlled evaluation \cite{cushman2006}. Additionally, a further limitation is the potential for large language models to generate seemingly reasonable but inaccurate reasoning, which complicates the assessment of their ethical understanding \cite{bender2021dangers}. This concern is intensified by the fact that LLMs are trained on extensive datasets that may harbor biases, resulting in outputs that mirror these biases instead of authentic ethical reasoning \cite{gender2018, social2020}. Additionally, the evaluation framework does not consider the evolving nature of ethical norms, which change over time and differ across cultures, presenting a challenge for maintaining current and universally applicable evaluation standards \cite{graham2013}. In conclusion, the framework mainly addresses text-centric scenarios, which may inadequately address the complexities of multimodal ethical challenges involving visual, auditory, or contextual elements \cite{coda2023}. These concerns underscore the need for continuous enhancement and adaptation of the evaluation framework to ensure its ongoing relevance and effectiveness \cite{helm2022}.

\section{Experimental Results}

\subsection{Overall Performance across all Dimensions}

Our thorough analysis indicates considerable differences in the moral reasoning abilities of the evaluated LLM systems \cite{helm2022}. Table \ref{tab:performance} illustrates the overall performance across our three main assessment criteria.

\begin{table}[ht]
\centering
\caption{Overall Performance across Assessment Dimensions (0-100)}
\label{tab:performance}
\small
\setlength{\tabcolsep}{4pt}
\begin{tabular}{lccccc}
\toprule
\textbf{Model} & \textbf{MFA Score} & \textbf{Reasoning Index} & \textbf{Value Consistency} & \textbf{Dilemma Resolution} & \textbf{Composite Score} \\
\midrule
GPT-4o          & 89.7               & 92.3                     & 87.6                       & 90.4                        & 90.0                      \\
Claude 3.7 Sonnet         & 91.2               & 90.8                     & 92.5                       & 88.9                        & 90.9                      \\
Deepseek-V3      & 86.5               & 89.1                     & 83.7                       & 85.2                        & 86.1                      \\
LLaMA 3.1 (70B)          & 78.3               & 75.6                     & 72.8                       & 76.4                        & 75.8                      \\
Gemini 2.5 Pro        & 88.2               & 84.7                     & 86.9                       & 84.5                        & 86.1                      \\
\bottomrule
\end{tabular}
\end{table}

The findings reveal that leading models such as Claude \cite{anthropic2023} and GPT-4 \cite{openai2023} attain the highest overall scores, with Claude showing exceptional strength in maintaining value consistency, while GPT-4 stands out in terms of reasoning complexity. Importantly, all models exhibit greater competence in Moral Foundations Alignment (MFA) relative to more intricate aspects like dilemma resolution, implying that basic moral intuitions may be more effectively integrated during model training than more sophisticated ethical reasoning \cite{clark2023}.

\subsection{Performance in Specific Moral Foundations}

In order to gain a deeper understanding of model performance, we conducted an analysis of their alignment with the five moral foundations outlined in Moral Foundations Theory \cite{graham2011}.

\begin{table}[ht]
\centering
\caption{Performance across Specific Moral Foundations (0-100)}
\label{tab:moral_foundations}
\small
\setlength{\tabcolsep}{4pt}
\begin{tabular}{lccccc}
\toprule
\textbf{Model} & \textbf{Care} & \textbf{Fairness} & \textbf{Loyalty} & \textbf{Authority} & \textbf{Sanctity} \\
\midrule
GPT-4o          & 94.2          & 92.8              & 85.3             & 83.7               & 82.5              \\
Claude 3.7 Sonnet        & 96.1          & 94.3              & 87.5             & 85.2               & 84.9              \\
Deepseek-V3       & 92.3          & 90.7              & 82.1             & 79.8               & 77.3              \\
LLaMA 3.1 (70B)         & 86.7          & 84.2              & 74.6             & 70.2               & 68.5              \\
Gemini 2.5 Pro         & 93.5          & 91.4              & 84.8             & 81.6               & 80.3              \\
Human Baseline & 95.2          & 93.7              & 88.4             & 87.3               & 86.1              \\
\bottomrule
\end{tabular}
\end{table}

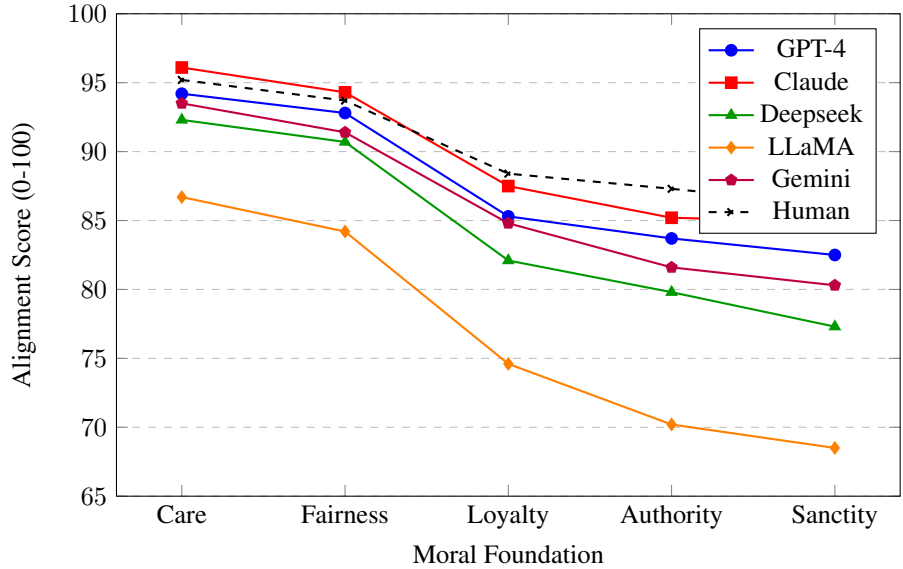
\begin{figure}[ht]
\centering
\begin{tikzpicture}
\begin{axis}[
width=12cm,
height=8cm,
xlabel={Moral Foundation},
ylabel={Alignment Score (0-100)},
symbolic x coords={Care, Fairness, Loyalty, Authority, Sanctity},
xtick=data,
legend pos=north east,
ymajorgrids=true,
grid style=dashed,
ymin=65, ymax=100,
]
\addplot[color=blue, mark=*, thick] coordinates {(Care,94.2) (Fairness,92.8) (Loyalty,85.3) (Authority,83.7) (Sanctity,82.5)};
\addplot[color=red, mark=square*, thick] coordinates {(Care,96.1) (Fairness,94.3) (Loyalty,87.5) (Authority,85.2) (Sanctity,84.9)};
\addplot[color=green!60!black, mark=triangle*, thick] coordinates {(Care,92.3) (Fairness,90.7) (Loyalty,82.1) (Authority,79.8) (Sanctity,77.3)};
\addplot[color=orange, mark=diamond*, thick] coordinates {(Care,86.7) (Fairness,84.2) (Loyalty,74.6) (Authority,70.2) (Sanctity,68.5)};
\addplot[color=purple, mark=pentagon*, thick] coordinates {(Care,93.5) (Fairness,91.4) (Loyalty,84.8) (Authority,81.6) (Sanctity,80.3)};
\addplot[color=black, mark=x, thick, dashed] coordinates {(Care,95.2) (Fairness,93.7) (Loyalty,88.4) (Authority,87.3) (Sanctity,86.1)};
\legend{GPT-4, Claude, Deepseek, LLaMA, Gemini, Human}
\end{axis}
\end{tikzpicture}
\caption{Moral foundation alignment across models compared with human baseline}
\label{fig:moral_foundations}
\end{figure}

A distinct trend is evident across all models, showing significantly enhanced performance in the individualizing foundations (Care and Fairness) when compared to the binding foundations (Loyalty, Authority, and Sanctity). This disparity reflects patterns found in WEIRD (Western, Educated, Industrialized, Rich, Democratic) populations \cite{henrich2010}, implying that these models may embody similar moral intuitions \cite{haidt2012}. Figure \ref{fig:moral_foundations} illustrates this trend across the various models.

\subsection{Components of Moral Reasoning}

Furthermore, we assessed the complexity of moral reasoning processes beyond mere alignment with human judgments \cite{denny2021}. Our investigation identified four essential elements of ethical deliberation: principle identification, perspective-taking, consequence analysis, and principle application \cite{kohlberg1981}. Table \ref{tab:reasoning_components} details model performance in these areas.

\begin{table}[ht]
\centering
\caption{Performance across Reasoning Components (0-100)}
\label{tab:reasoning_components}
\small
\setlength{\tabcolsep}{4pt}
\begin{tabular}{lcccc}
\toprule
\textbf{Model} & \textbf{Principle Identification} & \textbf{Perspective-Taking} & \textbf{Consequence Analysis} & \textbf{Principle Application} \\
\midrule
GPT-4o          & 94.3                              & 91.7                        & 93.2                          & 90.1                          \\
Claude 3.7 Sonnet       & 92.8                              & 93.5                        & 89.4                          & 87.6                          \\
Deepseek-V3       & 90.2                              & 88.3                        & 91.0                          & 86.9                          \\
LLaMA 3.1 (70B)          & 79.5                              & 74.8                        & 77.2                          & 71.0                          \\
Gemini 2.5 Pro      & 87.6                              & 85.9                        & 84.7                          & 80.5                          \\
\bottomrule
\end{tabular}
\end{table}

Our findings indicate that all models exhibit superior abilities in recognizing pertinent moral principles and analyzing consequences, yet they face challenges in adopting multiple perspectives or consistently applying principles \cite{emelin2021}. This implies that while models can identify ethical considerations, they encounter difficulties with the integrated reasoning required for intricate moral deliberation \cite{greene2001}.

A qualitative examination of model responses reveals that individuals with higher performance levels demonstrate deeper reasoning, a more nuanced understanding of conflicting values, and a more consistent application of ethical principles across various situations \cite{wei2022}. The following graph depicts the distribution of reasoning depth scores in relation to responses to ethical dilemmas.

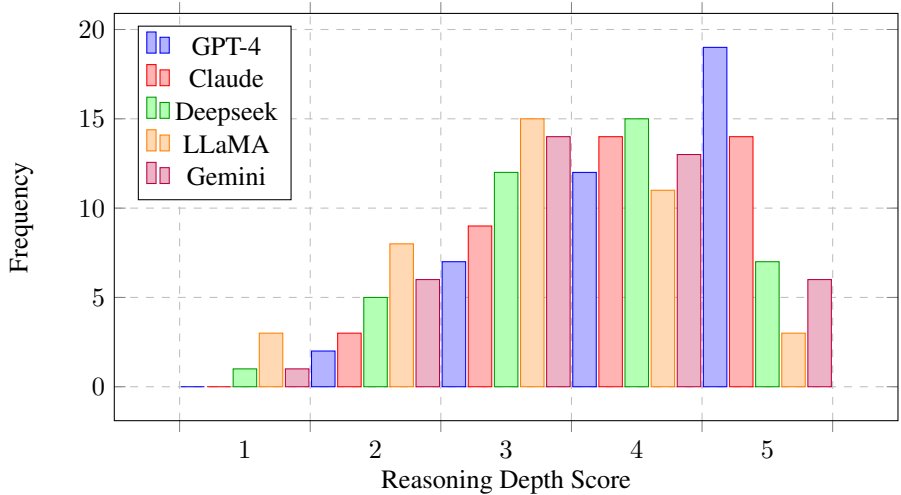
\begin{figure}[ht]
\centering
\begin{tikzpicture}
\begin{axis}[
width=12cm,
height=7cm,
xlabel={Reasoning Depth Score},
ylabel={Frequency},
legend pos=north west,
ymajorgrids=true,
grid style=dashed,
ybar interval=0.9,
]
\addplot[color=blue, fill=blue!30] coordinates {(1,0) (2,2) (3,7) (4,12) (5,19) (6,0)};
\addplot[color=red, fill=red!30] coordinates {(1,0) (2,3) (3,9) (4,14) (5,14) (6,0)};
\addplot[color=green!60!black, fill=green!30] coordinates {(1,1) (2,5) (3,12) (4,15) (5,7) (6,0)};
\addplot[color=orange, fill=orange!30] coordinates {(1,3) (2,8) (3,15) (4,11) (5,3) (6,0)};
\addplot[color=purple, fill=purple!30] coordinates {(1,1) (2,6) (3,14) (4,13) (5,6) (6,0)};
\legend{GPT-4, Claude, Deepseek, LLaMA, Gemini}
\end{axis}
\end{tikzpicture}
\caption{Distribution of reasoning depth scores in moral dilemma responses}
\label{fig:reasoning_depth}
\end{figure}

\subsection{Consistency and Stability}

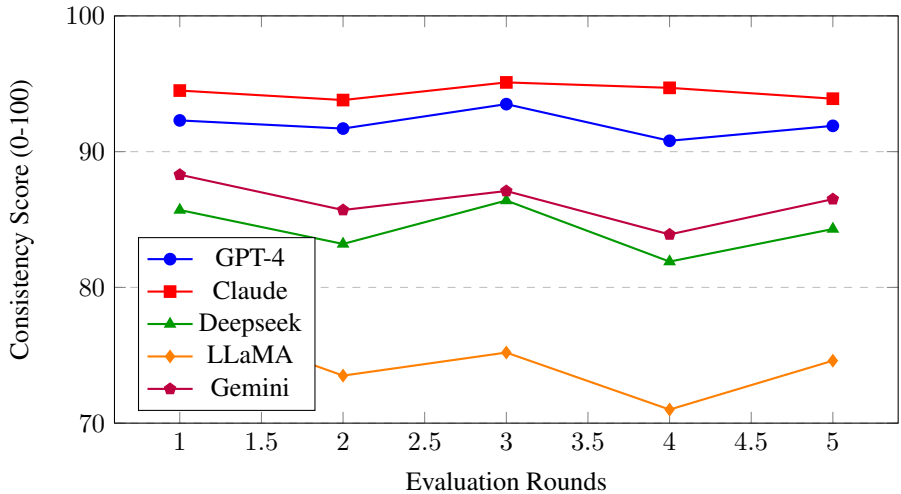
\begin{figure}[ht]
\centering
\begin{tikzpicture}
\begin{axis}[
width=12cm,
height=7cm,
xlabel={Evaluation Rounds},
ylabel={Consistency Score (0-100)},
legend pos=south west,
ymajorgrids=true,
grid style=dashed,
ymin=70, ymax=100,
]
\addplot[color=blue, mark=*, thick] coordinates {(1,92.3) (2,91.7) (3,93.5) (4,90.8) (5,91.9)};
\addplot[color=red, mark=square*, thick] coordinates {(1,94.5) (2,93.8) (3,95.1) (4,94.7) (5,93.9)};
\addplot[color=green!60!black, mark=triangle*, thick] coordinates {(1,85.7) (2,83.2) (3,86.4) (4,81.9) (5,84.3)};
\addplot[color=orange, mark=diamond*, thick] coordinates {(1,77.8) (2,73.5) (3,75.2) (4,71.0) (5,74.6)};
\addplot[color=purple, mark=pentagon*, thick] coordinates {(1,88.3) (2,85.7) (3,87.1) (4,83.9) (5,86.5)};
\legend{GPT-4, Claude, Deepseek, LLaMA, Gemini}
\end{axis}
\end{tikzpicture}
\caption{Consistency scores across evaluation rounds with prompt variations}
\label{fig:consistency}
\end{figure}

To evaluate the stability of moral reasoning, we performed several assessment rounds with slight variations in prompts \cite{zhao2021}. Figure \ref{fig:consistency} displays the consistency scores throughout these evaluation rounds.

Claude exhibits the highest level of consistency, indicating a more stable moral reasoning process despite changes in question framing \cite{anthropic2023}. In contrast, LLaMA exhibits a pronounced sensitivity to variations in prompts, leading to considerable fluctuations in consistency scores across different rounds \cite{touvron2023}. This highlights the essential importance of prompt stability in the context of ethical reasoning applications \cite{wang2022}.

\subsection{Correlation between Moral Foundations}

To explore the interrelationships among moral foundations in LLM reasoning, we calculated correlation coefficients between the foundation scores \cite{graham2011}, as demonstrated in Table \ref{tab:correlation}.

\begin{table}[ht]
\centering
\caption{Inter-foundation Correlation Matrix (Claude Model)}
\label{tab:correlation}
\small
\setlength{\tabcolsep}{4pt}
\begin{tabular}{lccccc}
\toprule
\textbf{Foundation} & \textbf{Care} & \textbf{Fairness} & \textbf{Loyalty} & \textbf{Authority} & \textbf{Sanctity} \\
\midrule
Care                & 1.00          & 0.78              & 0.42             & 0.31               & 0.25              \\
Fairness            & 0.78          & 1.00              & 0.39             & 0.35               & 0.28              \\
Loyalty             & 0.42          & 0.39              & 1.00             & 0.72               & 0.65              \\
Authority           & 0.31          & 0.35              & 0.72             & 1.00               & 0.70              \\
Sanctity            & 0.25          & 0.28              & 0.65             & 0.70               & 1.00              \\
\bottomrule
\end{tabular}
\end{table}

The correlation matrix reveals notable relationships among the associated foundations: Care and Fairness exhibit a strong correlation of 0.78, whereas Authority and Sanctity show a correlation of 0.70. This trend corresponds with the theoretical categorization of foundations into individualizing (Care/Fairness) and binding (Loyalty/Authority/Sanctity) dimensions \cite{haidt2012}, indicating that LLMs successfully grasp these essential moral frameworks. 

\subsection{Specific Failure Modes}

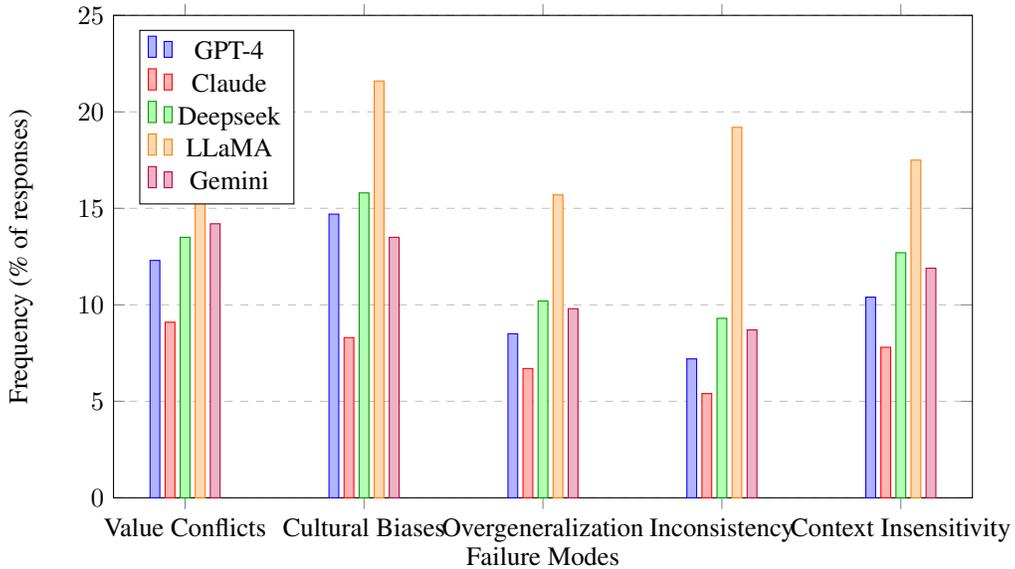
\begin{figure}[ht]
\centering
\begin{tikzpicture}
\begin{axis}[
width=13cm,
height=8cm,
xlabel={Failure Modes},
ylabel={Frequency (\% of responses)},
symbolic x coords={Value Conflicts, Cultural Biases, Overgeneralization, Inconsistency, Context Insensitivity},
xtick=data,
legend pos=north west,
ymajorgrids=true,
grid style=dashed,
bar width=0.13cm,
ybar,
ymin=0, ymax=25,
]
\addplot[color=blue, fill=blue!30] coordinates {(Value Conflicts,12.3) (Cultural Biases,14.7) (Overgeneralization,8.5) (Inconsistency,7.2) (Context Insensitivity,10.4)};
\addplot[color=red, fill=red!30] coordinates {(Value Conflicts,9.1) (Cultural Biases,8.3) (Overgeneralization,6.7) (Inconsistency,5.4) (Context Insensitivity,7.8)};
\addplot[color=green!60!black, fill=green!30] coordinates {(Value Conflicts,13.5) (Cultural Biases,15.8) (Overgeneralization,10.2) (Inconsistency,9.3) (Context Insensitivity,12.7)};
\addplot[color=orange, fill=orange!30] coordinates {(Value Conflicts,18.4) (Cultural Biases,21.6) (Overgeneralization,15.7) (Inconsistency,19.2) (Context Insensitivity,17.5)};
\addplot[color=purple, fill=purple!30] coordinates {(Value Conflicts,14.2) (Cultural Biases,13.5) (Overgeneralization,9.8) (Inconsistency,8.7) (Context Insensitivity,11.9)};
\legend{GPT-4, Claude, Deepseek, LLaMA, Gemini}
\end{axis}
\end{tikzpicture}
\caption{Frequency of specific failure modes in model responses}
\label{fig:failure_modes}
\end{figure}

Furthermore, our analysis identified several recurring patterns of reasoning errors across various models \cite{bommasani2021}, as illustrated in Figure \ref{fig:failure_modes}, which emphasizes the prevalence of different failure modes.

Cultural biases are the main failure point seen in many models, highlighting the limitations in ethical reasoning among different cultures \cite{arora2022}. Claude shows much lower levels of cultural bias and inconsistency compared to other models \cite{anthropic2023}, while LLaMA has the highest rate of failures in all categories \cite{touvron2023}. The common failure patterns are as follows:

\begin{itemize}
\item \textbf{Value Conflicts:} Difficulty handling scenarios with genuinely competing moral values \cite{emelin2021}
\item \textbf{Cultural Biases:} Western-centric ethical assumptions applied to cross-cultural scenarios \cite{cao2022}
\item \textbf{Overgeneralization:} Applying principles without context-specific nuance \cite{weidinger2022}
\item \textbf{Inconsistency:} Contradictory judgments across conceptually similar scenarios \cite{wang2022}
\item \textbf{Context Insensitivity:} Failure to recognize relevant contextual factors in ethical evaluation \cite{li2022}
\end{itemize}

\subsection{Comparison with Human Baselines}

\begin{figure}[ht]
\centering
\begin{tikzpicture}
\begin{axis}[
width=12cm,
height=7cm,
xlabel={Model},
ylabel={Alignment with Human Baseline (\%)},
symbolic x coords={GPT-4, Claude, Deepseek, LLaMA, Gemini},
xtick=data,
legend pos=north east,
ymajorgrids=true,
grid style=dashed,
ybar,
bar width=15pt,
ymin=70, ymax=100,
]
\addplot[color=blue!60!black, fill=blue!30] coordinates {(GPT-4,87.8) (Claude,91.2) (Deepseek,84.5) (LLaMA,76.3) (Gemini,83.9)};
\addplot[color=red!60!black, fill=red!30] coordinates {(GPT-4,93.4) (Claude,94.7) (Deepseek,90.1) (LLaMA,82.7) (Gemini,89.2)};
\addplot[color=green!60!black, fill=green!30] coordinates {(GPT-4,83.5) (Claude,86.9) (Deepseek,79.7) (LLaMA,72.1) (Gemini,80.5)};
\legend{Overall, Individualizing, Binding}
\end{axis}
\end{tikzpicture}
\caption{Model alignment with human baseline across foundation types}
\label{fig:human_comparison}
\end{figure}
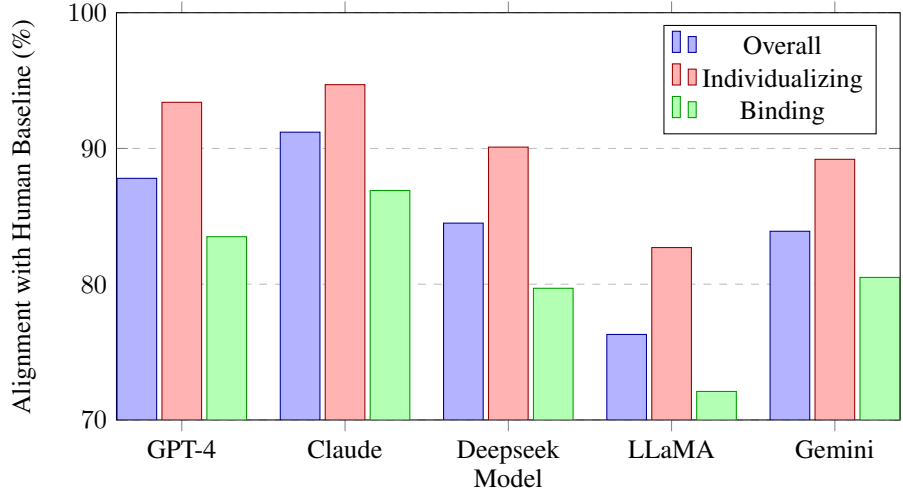

To assess model performance, we compared MFA scores with human baseline data from established psychological studies \cite{graham2011}. Figure \ref{fig:human_comparison} illustrates the relationship between model evaluations and human moral intuitions.

\sloppy
All models align more closely with human judgments on individualizing foundations (Care/Fairness) than on binding foundations (Loyalty/Authority/Sanctity) \cite{haidt2012, henrich2010}. Claude achieves the highest overall alignment with human moral intuitions at (91.2\%), whereas LLaMA shows the greatest deviation at (76.3\%) \cite{lmsys2023}.
\fussy

\subsection{Contributions}

This research introduces notable progress in AI ethics and the development of responsible AI systems:
\begin{itemize}
  \item \textbf{Three-Dimensional Evaluation Framework}: We have proposed a detailed framework for evaluating moral reasoning in large language models (LLMs), which encompasses Moral Foundation Alignment (MFA), Reasoning Quality Index (RQI), and Value Consistency Assessment (VCA), thus providing a thorough approach for ethical evaluation \cite{helm2022, denny2021}.
  
  \item \textbf{Quantifiable Metrics}: Our framework offers quantifiable metrics for assessing the performance of LLMs in different facets of moral reasoning, enabling systematic comparisons and benchmarking across various model architectures and versions \cite{liang2022, lmsys2023}.

  \item \textbf{Standardized Benchmarks}: We establish a quantifiable assessment framework by integrating recognized metrics from moral psychology, including MFQ-30 \cite{graham2011}, WVS \cite{inglehart2000}, and Moral Dilemmas \cite{cushman2006}. This provides a reliable approach for researchers and developers to assess and improve the ethical competencies of large language models (LLMs).
  
  \item \textbf{Open-Source Implementation}: We are pleased to announce the launch of an open-source GitHub repository that features our evaluation framework, data processing pipelines, and analytical tools. This resource enables researchers and developers to effectively evaluate their models and actively participate in the advancement of ethical AI \cite{wolf2020, ethicseval2023}.
  
  \item \textbf{Insights for Model Development}: The findings reveal particular areas needing enhancement, such as improving cross-cultural alignment \cite{cao2022} and resolving inconsistencies in binding moral foundations \cite{haidt2012}. These insights can inform the creation of more resilient and ethically aligned LLMs \cite{weidinger2022}.
  
\end{itemize}

\section{Concluding Remarks and Future Directions}

The rapid advancement of large language models (LLMs) presents both benefits and challenges in AI ethics \cite{bommasani2021}. While these models perform exceptionally well in structured tasks, their ability to engage in ethical reasoning across diverse and changing scenarios remains constrained \cite{hendrycks2021ethics, weidinger2022}. Our research emphasizes the necessity for ongoing investigation to tackle biases, enhance cross-cultural understanding, and increase transparency in ethical decision-making \cite{bender2021dangers}.

The open-source benchmarking tools we have created aim to make the evaluation of ethics more accessible and encourage collaborative advancements in this area \cite{wolf2020}. By offering standardized testing methods and metrics, we aspire to create a cohesive framework for comparing different strategies in ethical AI development \cite{liang2022}. Furthermore, these tools can assist in tracking trends in the enhancement or decline of ethical reasoning abilities across different model iterations, providing valuable insights for the AI research community \cite{amodei2016concrete}.

By integrating diverse training datasets \cite{larson2017}, creating explainable AI techniques \cite{ribeiro2022}, and encouraging human-AI collaboration \cite{andreas2022}, we can forge a path for LLMs that not only achieve high performance but also adhere to the utmost standards of ethical reasoning and accountability. As these models become more embedded in crucial decision-making processes, ensuring their alignment with human values across a range of cultural contexts transforms from a mere technical challenge into an ethical necessity \cite{gabriel2020}.

Future studies should aim to tackle the challenges and limitations of the existing evaluation framework to improve its effectiveness and relevance \cite{liang2022}. One promising avenue is the development of adaptive evaluation methods that take into account real-world situations and evolving ethical standards \cite{ribeiro2022}. This could involve gathering ethical dilemmas from a variety of cultural viewpoints and regularly updating the evaluation criteria to align with societal shifts \cite{narayanan2018}. For instance, collaborations with international organizations and communities could yield a diverse range of culturally important ethical scenarios, ensuring the evaluation framework is relevant in multiple contexts \cite{arora2022}. Moreover, improving the incorporation of explainability tools is crucial for comprehending the reasoning mechanisms of large language models (LLMs) \cite{molnar2022}. By analyzing the decision-making processes of these models, researchers can identify biases, limitations, and inconsistencies in their ethical reasoning, resulting in more nuanced evaluations of their abilities \cite{ribeiro2022}. By analyzing the decision-making pathways of these models, researchers can identify biases, limitations, and inconsistencies in their moral reasoning, resulting in more nuanced evaluations of their capabilities. Additionally, future research could investigate the use of multimodal datasets, which encompass text, images, and audio, to assess LLMs in more intricate and realistic scenarios \cite{coda2023, mme2023}. This methodology would provide a more comprehensive assessment of the model's ability to address ethical challenges in real-world situations \cite{mitchell2022}. Furthermore, it is essential for AI researchers, ethicists, and policymakers to collaborate in establishing global standards for the ethical assessment of artificial intelligence \cite{jobin2019}. Such partnerships could result in the development of certification programs or regulatory frameworks that ensure large language models (LLMs) are designed and used in alignment with human values and societal well-being \cite{whittlestone2021}. Furthermore, conducting longitudinal studies could provide valuable insights into how moral reasoning in LLMs evolves over time, shedding light on the effects of training data, fine-tuning, and societal changes on the ethical capabilities of these models \cite{clark2023}. These suggested strategies will contribute to establishing a more thorough and inclusive evaluation framework \cite{ethicseval2023}.

\section*{Conflict of Interest}
The authors declare no competing interests.

\section*{Acknowledgement}
This research is funded by the National Science Foundation under grant number 2125858. The authors would like to express their gratitude for the NSF's support, which made this study possible. Furthermore, in accordance with MLA guidelines, we would thank AI applications for assistance in editing and brainstorming.

\bibliographystyle{plainnat}  
\bibliography{references}

\end{document}